\documentclass[12pt,showpacs,preprintnumbers,amsmath,amssymb]{revtex4}
\usepackage{graphicx}
\begin{document}

\newcommand{\pderiv}[2]{\frac{\partial #1}{\partial #2}}
\newcommand{\deriv}[2]{\frac{d #1}{d #2}}

\title{Consequences of the H-Theorem from Nonlinear Fokker-Planck Equations}

\vskip \baselineskip

\author{Veit Schw\"ammle}
\thanks{E-mail address: veit@cbpf.br}

\author{Fernando D. Nobre}
\thanks{E-mail address: fdnobre@cbpf.br}

\author{Evaldo M.~F. Curado}
\thanks{E-mail address: evaldo@cbpf.br}

\address{
Centro Brasileiro de Pesquisas F\'{\i}sicas \\
Rua Xavier Sigaud 150 \\
22290-180 \hspace{5mm} Rio de Janeiro - RJ \hspace{5mm} Brazil}

\date{\today}


\begin{abstract}
\noindent
A general type of nonlinear Fokker-Planck equation is derived directly from
a master equation, by introducing generalized transition rates. 
The H-theorem is demonstrated for systems that 
follow those classes of 
nonlinear Fokker-Planck equations, in the presence of an external
potential. For that, a relation involving terms 
of Fokker-Planck equations and general entropic forms is
proposed. It is shown that, at equilibrium, this relation is equivalent to
the maximum-entropy principle. Families of 
Fokker-Planck equations may be related to a single type of entropy, and so,
the correspondence between well-known entropic forms and their
associated Fokker-Planck equations is explored. It is shown that 
the Boltzmann-Gibbs
entropy, apart from its connection with the standard -- linear Fokker-Planck
equation -- may be also related to a family of nonlinear 
Fokker-Planck equations. 

\vskip \baselineskip


\end{abstract}
\maketitle

\newpage

\noindent
{\large\bf 1. \quad Introduction}

\vskip \baselineskip

It is well known that many real systems exhibit a dynamical behavior that
falls out of the scope of the  
standard linear differential equations of physics. Although the linear
Fokker-Planck equation (FPE) \cite{vankampen} is considered appropriate
for the  
description of a wide variety of physical phenomena -- typically those
associated with normal diffusion -- it is well accepted that this equation
is not adequate for describing anomalous diffusion. 
An example consists in
particle transport in disordered media \cite{muskat}, like amorphous materials, or some
other kind of media containing impurities and/or defects. In such systems,
particles are driven by highly irregular forces, which lead to transport
coefficients that may vary locally in a nontrivial manner. Among other phenomena that also fall out of the scope of the linear FPE, one may mention 
surface growth \cite{spohn}, diffusion of micelles in salted water 
\cite{bouchaud91}, and the heart-beat histograms in a healthy
individual \cite{peng}. 
These nonlinear phenomena became one of the most investigated topics in
physics nowadays, with a lot of applications in natural systems. Many
interesting new aspects appear, leading to a wide range of open
problems in physics. 

In order to cope with such anomalous systems, modifications in the linear 
FPE have been carried out, and this subject has attracted the attention of many 
researchers recently. Essentially,
there are two alternatives for introducing modifications in the linear FPE:
(i) a procedure that leads to the fractional FPE 
(see ref. \cite{metzler00} for a review), where one considers a linear 
theory with nonlocal operators carrying the anomalous nature 
of the process; (ii) the nonlinear FPEs \cite{frankbook}, 
that in most of the cases come out as simple 
phenomenological generalizations of the standard linear FPE
\cite{plastino95,tsallis96,borland98,%
borland99,lenzi03,frank99,frank01a,malacarne01,malacarne02,
chavanis03,chavanis04}. 
Recently, it has been shown that nonlinear FPEs may be derived directly
from a standard master equation, by 
introducing nonlinear effects on its associated transition probabilities
\cite{curado03,nobre04,veit07a}. 

The nonextensive statistical 
mechanics formalism has emerged naturally as a strong candidate for dealing
appropriately with many real systems that are not satisfactorily described
within  standard (extensive) statistical mechanics
\cite{next99,next04,next05}.
The power-like probability distribution that maximizes the entropy 
proposed by Tsallis \cite{tsallis88,curado91,tsallis98} is very often found
as solution of nonlinear FPEs \cite{plastino95,tsallis96,borland98,borland99,
frank01a,lenzi03,chavanis03,chavanis04}, suggesting that the nonextensive 
statistical mechanics formalism should be 
intimately related to nonlinear FPEs. 

Many important equations and properties of standard statistical mechanics 
have been extended within the
formalism of nonextensive statistical mechanics. An example is the 
H-theorem, which was shown to be valid, taking into account certain 
restrictions on the 
parameters of the corresponding entropic form 
\cite{frankbook,frank01a,mariz92,ramshaw93,shiino01}. Usually, one proves
the H-theorem by defining previously an entropic form, and then considering
either the master equation or a FPE, when dealing with the time derivative 
of the probability distribution.

The main motivation of this paper is to prove the H-theorem for a system in
the presence of an external potential and following a general type of
nonlinear FPE. 
In order to achieve this, we introduce a relation involving quantities of
the FPE with an entropic form; in principle, one may have classes of
Fokker-Planck equations associated with a single entropic form. 
We show that, when considered at
equilibrium, this relation is equivalent to the maximum-entropy principle. 
In the next section we derive a general FPE, directly from a master
equation, by introducing nonlinear terms in its transition probabilities;
such a FPE will be used throughout most of this paper. 
In section 3 we prove the H-theorem by using this FPE, 
and show that the validity of this
theorem requires a relation involving a general entropic form
and the parameters of this nonlinear FPE. 
In section 4 we discuss particular
cases of this FPE and their associated entropic forms.
In section 5 we introduce a modified FPE that is compatible with the
definition of a  
``generalized internal energy'', as used within the context of 
nonextensive statistical mechanics. 
The same relation introduced previously is also necessary in this case, in
order to prove the H-theorem.
Finally, in section 6 we present our conclusions. 

\vskip 2 \baselineskip
\noindent
{\large\bf 2. \quad Derivation of the Nonlinear Fokker-Planck Equation from
a Master Equation}

\vskip \baselineskip

In this section we will derive, directly from the master equation, the
nonlinear FPE that will be investigated throughout most of the 
present paper; we will follow closely the approach used in 
Refs. \cite{curado03,nobre04}. Let us then consider the standard master
equation, associated with a discrete spectrum, 

$$
\frac{\partial P(n,t)}{\partial t} = \sum \limits_{m=-\infty}^{\infty} 
\left[ P(m,t)w_{m,n}(t) - P(n,t)w_{n,m}(t) \right]~,
\eqno(2.1)
$$

\vskip \baselineskip
\noindent
with $P(n,t)$ representing the probability  
for finding a given system in a state characterized by a variable $n$, at
time $t$. We introduce nonlinearities in the system through the following
transition rates,

$$
w_{k,l}(\Delta) = - \frac{1}{\Delta} \delta_{k,l+1} A(k\Delta)
~a[P(k\Delta,t)] +
\frac{1}{\Delta^2} \left( \delta_{k,l+1} + \delta_{k,l-1} \right)
~\Upsilon[P(k\Delta,t),R(l\Delta,t)]~.  
\eqno(2.2)
$$

\vskip \baselineskip
\noindent
In the equation above, $A(k\Delta)$ represents an external dimensionless
force, $a[P]$ is a functional of the probability $P(n,t)$, whereas
the functional $\Upsilon[P,R]$ depends on two probabilities $P$ and $R$
that are associated to two different states, although 
$R(k\Delta,t) \equiv P(k\Delta,t)$. Substituting
this transition rate in Eq. (2.1), performing the sums, and defining
$x=k\Delta$, one gets

\setcounter{enumi}{2}
\setcounter{enumii}{3}
\renewcommand{\theequation}{\arabic{enumi}.\arabic{enumii}}
\begin{eqnarray}
  \lefteqn{ \frac{\partial P(x,t)}{\partial t}   =} \nonumber \\
& &   -\frac{1}{\Delta} \Bigl\{ P(x+\Delta,t)
    A(x+\Delta)~a [P(x+\Delta,t)] - P(x,t)A(x)~a [P(x,t)] 
    \Bigr\}  \nonumber \\
& &     + \frac{1}{\Delta^2} \Bigl\{ P(x+\Delta,t)~
  \Upsilon[P(x+\Delta,t),R(x,t)] 
  + P(x-\Delta,t)~\Upsilon[P(x-\Delta,t),R(x,t)]
\Bigr\} \nonumber\\
& & - \frac{1}{\Delta^2} \ P(x,t) \Bigl\{
        \Upsilon[P(x,t),R(x+\Delta,t)] 
	+ \Upsilon[P(x,t),R(x-\Delta,t)] \Bigr\}~. 
\end{eqnarray}

\vspace{5mm}

\noindent
The quantities depending on $\Delta$ may be expanded for small $\Delta$, e.g., 

\setcounter{enumi}{2}
\setcounter{enumii}{4}
\renewcommand{\theequation}{\arabic{enumi}.\arabic{enumii}}
\begin{eqnarray}
\Upsilon [P(x,t), R(x \pm \Delta,t)] &  = & \left[ 
\Upsilon [P(x,t), R(x,t)] + \left(
\pm \Delta {\partial R(x,t) \over \partial x} + {\Delta^{2} \over 2}
{\partial ^{2} R(x,t) \over \partial x^{2}} \right) 
{\partial \Upsilon [P,R] \over \partial R} \right. 
\nonumber \\ \nonumber \\
& & \left. 
+ \ {\Delta^{2} \over 2} \left( {\partial R(x,t) \over \partial x} \right)^{2}
{\partial ^{2} \Upsilon [P,R] \over \partial R^{2}} + \cdots \right]_{R=P},   
\end{eqnarray}

\vspace{5mm}

\noindent
in such a way that considering the limit $\Delta \rightarrow 0$, one gets
the nonlinear FPE,  

$$
{\partial P(x,t) \over \partial t}  = 
-{\partial \{A(x) \Psi [P(x,t)]\} \over \partial x} 
+{\partial \over \partial x} 
\left\{\Omega [P(x,t)]{\partial P(x,t) \over \partial x}\right\}~,
\eqno(2.5)
$$

\vspace{5mm}

\noindent
with 

\setcounter{enumi}{2}
\setcounter{enumii}{6}
\setcounter{equation}{0}
\renewcommand{\theequation}{\arabic{enumi}.\arabic{enumii}\alph{equation}}
\begin{eqnarray}
\Psi [P(x,t)] & = & P(x,t)a [P(x,t)]~,  
\\ \nonumber \\  
\Omega[P(x,t)] & = & \left[ \Upsilon [P,R] + P(x,t) \left( 
{\partial \Upsilon [P,R] \over \partial P}
- {\partial \Upsilon [P,R] \over \partial R} \right) \right]_{R=P},   
\end{eqnarray}

\vspace{5mm}

\noindent  
where we have used the fact that 
$\partial P(x,t) / \partial x \equiv \partial R(x,t) / \partial x$.
The external force $A(x)$ is associated with a potential $\phi(x)$
[$A(x)=-d\phi(x)/dx$,  $\phi(x)=-\int_{-\infty}^{x}A(x^{\prime})dx^{\prime}$], 
and we are 
assuming analyticity of the potential $\phi(x)$, as well as integrability 
of the force $A(x)$ in all space. 
Furthermore, the functionals $\Psi [P(x,t)]$ 
and $\Omega [P(x,t)]$ are supposed to be both positive finite quantities,
integrable, as well as   
differentiable (at least once) with respect to the 
probability distribution $P(x,t)$, i.e., they should be at least,  
$\Omega[P], \Psi[P] \in C^{1}$. In addition to that, 
$\Psi [P(x,t)]$ should be also a monotonically increasing functional 
of $P(x,t)$. 

As usual, we assume that the probability
distribution, together with its first derivative, as well as the product   
$A(x)\Psi[P(x,t)]$, should all be zero at infinity, 

$$
P(x,t)|_{x \rightarrow \pm \infty} = 0~; \quad
\left. {\partial P(x,t) \over \partial x} 
\right|_{x \rightarrow \pm \infty} = 0~; \quad 
A(x)\Psi[P(x,t)]|_{x \rightarrow \pm \infty} = 0 \quad  
(\forall t)~.  
\eqno(2.7)
$$

\vskip \baselineskip
\noindent
The conditions above guarantee the preservation of the normalization for the
probability distribution, i.e., if for a given time $t_{0}$ one has
that 
$\int_{-\infty}^{\infty}dx \ P(x,t_{0}) = 1$, then a simple integration of 
Eq.~(2.5) with respect to the variable $x$ yields,  

$$
\frac{\partial }{\partial t} \int_{-\infty}^{\infty}dx \ P(x,t) = 
- \left[ A(x)\Psi[P(x,t)] \right]_{-\infty}^{\infty} 
+\left( \Omega[P(x,t)] \pderiv{P(x,t)}{x} \right)_{-\infty}^{\infty} = 0~, 
\eqno(2.8)
$$

\vskip \baselineskip
\noindent
and so, 

$$
\int_{-\infty}^{\infty}dx \ P(x,t) = 
\int_{-\infty}^{\infty}dx \ P(x,t_{0}) = 1 \qquad (\forall t)~. 
\eqno(2.9)
$$

\vskip \baselineskip
\noindent
It is important to stress that the nonlinear FPE of Eq. (2.5) is very
general and  
reproduces well-known FPEs of the literature. As examples, one should
mention  
the particular cases: (i) the linear FPE is recovered for   
$\Psi [P(x,t)]=P(x,t)$ and  
$\Omega=D$ (constant); (ii) the nonlinear FPE that presents 
Tsallis distribution as a solution \cite{plastino95,tsallis96}, 
is obtained by setting 
$\Psi [P(x,t)]=P(x,t)$ and $\Omega [P(x,t)]=qD[P(x,t)]^{q-1}$, where $q$ is 
the well-known entropic index \cite{tsallis88}, characteristic of the 
nonextensive statistical mechanics formalism; (iii) the nonlinear FPE 
derived previously from the master equation
\cite{curado03,nobre04,veit07a} is recovered for $\Psi [P(x,t)]=P(x,t)$. 

In the next section we prove the H-theorem for a system in
the presence of an external potential and following the general type of
nonlinear FPE of Eq. (2.5). 
   
\vskip 2 \baselineskip
\noindent
{\large\bf 3. \quad The H-Theorem}

\vskip \baselineskip

Herein, we will consider a general type of entropic form, 
satisfying the following conditions,

$$
S[P] = \Lambda (Q[P])~; \quad Q[P] = \int_{-\infty}^{\infty}dx \ g[P(x,t)]~;  
\quad g(0)=g(1)=0~;  \quad  
\frac{d^{2}g}{dP^{2}} \leq 0~,
\eqno(3.1)
$$

\vskip \baselineskip
\noindent
where $\Lambda[Q]$ represents a monotonically increasing outer functional
with dimensions 
of entropy that is supposed to satisfy, at least,  
$\Lambda[Q] \in C^{1}$, 
whereas the inner functional $g[P(x,t)]$ should be also, at least,
$g[P(x,t)] \in C^{2}$ in the interval  
$0<P(x,t)<1$ (end points excluded). 
Since we are dealing with a system that exchanges energy with its
surrounding, herein represented by the potential $\phi(x)$, it is important
to define also the free-energy functional, 

$$
F = U - {1 \over \beta} \ S~; \qquad U =  \int_{-\infty}^{\infty}dx \ \phi(x) 
P(x,t)~,  \eqno(3.2)
$$

\vskip \baselineskip
\noindent
where $\beta$ represents a positive Lagrange multiplier.   

The H-theorem, for a system subject to an external potential, corresponds
to a well-defined sign for the time derivative of the above free-energy
functional, which we will consider as $(dF / dt) \leq 0$.
Using the definitions above, 
  
\vspace{-5mm}
\setcounter{enumi}{3}
\setcounter{enumii}{3}
\renewcommand{\theequation}{\arabic{enumi}.\arabic{enumii}}
\begin{eqnarray}
\deriv{F}{t} & = & \pderiv{}{t} \left(\int_{-\infty}^{\infty}dx \ \phi(x) P(x,t)
- \frac{1}{\beta} \Lambda(Q[P])  \right)  \nonumber \\ 
\nonumber \\
& = & \int_{-\infty}^{\infty}dx \ \left( \phi(x)  
- \frac{1}{\beta} \deriv{\Lambda[Q]}{Q} \deriv{g[P]}{P} \right) \pderiv{P}{t}~, 
\end{eqnarray}

\vskip \baselineskip
\noindent
where we remind that $\Lambda[Q]$ and 
$d\Lambda[Q] / dQ$ do not depend on the variable $x$.  
Now, one may use the FPE of Eq. (2.5) for the time derivative of the
probability distribution; carrying an integration by parts, and assuming
the conditions of Eq. (2.7), one gets,  

\vspace{-5mm}
\setcounter{enumi}{3}
\setcounter{enumii}{4}
\begin{eqnarray}
\deriv{F}{t} & = & - \int_{-\infty}^{\infty}dx \ \left( 
\deriv{\phi(x)}{x}\Psi[P] 
+\Omega[P] \pderiv{P}{x} \right) \nonumber \\ 
\nonumber \\
& \times &
\left( \deriv{\phi(x)}{x} - \frac{1}{\beta} 
\deriv{\Lambda[Q]}{Q} \deriv{^2 g[P]}{P^2} \pderiv{P}{x} \right)~.
\end{eqnarray}

\vskip \baselineskip
\noindent
In most of the cases, one is interested in verifying the H-theorem by using 
a well-defined FPE,
together with particular entropic forms, 
in such a way that some of the quantities, $\Lambda[Q]$, $\Omega[P]$, 
$\Psi[P]$, 
and $d^{2}g[P]/d P^{2}$, are previously defined (see, e.g., 
Refs. \cite{frank01a,shiino01}). Herein, we follow a more general 
approach, i.e., we assume that Eqs. (2.5), (2.7), (3.1), and (3.2) are
satisfied, and then, we impose the condition

$$
- \frac{1}{\beta} \deriv{\Lambda[Q]}{Q} \deriv{^2 g[P]}{P^2} =  
\frac{\Omega[P]}{\Psi[P]}~. 
\eqno(3.5)
$$

\vskip \baselineskip
\noindent
Using this condition, Eq. (3.4) may be written as

$$
\deriv{F}{t} = - \int_{-\infty}^{\infty}dx \ \Psi[P]   
\left( \deriv{\phi(x)}{x} +
\frac{\Omega[P]}{\Psi[P]} \pderiv{P}{x} \right)^{2} \leq 0~,  
\eqno(3.6)
$$

\vskip \baselineskip
\noindent
and we remind that $\Psi[P]$ is a positive, monotonically 
increasing functional of $P(x,t)$.

It should be stressed that Eq. (3.5) expresses an important 
relation involving quantities of the FPE and possible entropic forms, 
for the case of a system in the presence of an external potential. 
It leads to a correspondence between whole families of 
FPEs, defined in terms of the
functionals $\Omega[P]$ and $\Psi[P]$, with a single
entropic form. 
Therefore, it allows the calculation of the entropic form associated with a
given class of FPEs; on the other hand, 
one may also start by considering a given 
entropic form and then find the class of FPEs associated to it. 
In fact, since the FPE is 
a phenomenological equation that specifies the dynamical
evolution associated with a given physical system, 
Eq.~(3.5) may be useful in the identification of  
the entropic form associated with such a system. In particular, one may
identify entropic forms 
associated with some anomalous systems, exhibiting unusual behavior that
are appropriately described by nonlinear FPEs, like the one of 
Eq.~(2.5). Within the present approach,   
the relation of Eq.~(3.5) should hold for the H-theorem to be valid;  
even though the
relation of Eq. (3.5) may not be unique, we shall argue its
relevance in what follows. 

First of all, let us show that at equilibrium, Eq. (3.5) is equivalent to the
maximum-entropy principle. For that, we introduce the functional 

$$
{\cal I}[P(x,t)] = \Lambda (Q[P])
+ \alpha \left( 1 - \int_{-\infty}^{\infty}dx \ P(x,t) \right)
+ \beta \left(U - \int_{-\infty}^{\infty}dx \ \phi(x) P(x,t) \right)~,
\eqno(3.7)
$$

\vskip \baselineskip
\noindent 
where $\alpha$ and $\beta$ are Lagrange multipliers. Then, one has that, 

$$
\left. \deriv{{\cal I}[P]}{P} 
\right|_{P = P_{\rm eq}(x)}= 0 \quad \Rightarrow \quad 
\left.  \deriv{\Lambda[Q]}{Q} \deriv{g[P]}{P}\right|_{P = P_{\rm eq}(x)} 
- \alpha - \beta \, \phi(x) = 0~,  
\eqno(3.8)
$$

\vskip \baselineskip
\noindent 
where $P_{\rm eq}(x)$ represents the equilibrium probability distribution. 

>From the general FPE of Eq. (2.5), one gets that, at equilibrium,

$$
A(x) = {\Omega[P_{\rm eq}] \over \Psi[P_{\rm eq}]} 
\deriv{P_{\rm eq}(x)}{x}~,  
\eqno(3.9)
$$

\vskip \baselineskip
\noindent
which, after integration, yields,

$$
\phi_{0}-\phi(x) = \int_{x_{0}}^{x}dx \  
{\Omega[P_{\rm eq}] \over \Psi[P_{\rm eq}]} 
\deriv{P_{\rm eq}(x)}{x}
= \int_{P_{\rm eq}(x_{0})}^{P_{\rm eq}(x)} \ 
{\Omega[P_{\rm eq}(x^{\prime})] \over 
\Psi[P_{\rm eq}(x^{\prime})]} 
dP_{\rm eq}(x^{\prime})~,  
\eqno(3.10)
$$

\vskip \baselineskip
\noindent
where $\phi_{0} \equiv \phi(x_{0})$ is a constant. 
Integrating Eq. (3.5), at equilibrium, 

$$
\frac{1}{\beta} \left. \deriv{\Lambda[Q]}{Q} 
\deriv{g[P]}{P}\right|_{P_{\rm eq}(x)} 
 = \phi(x) + C_{1}~, 
\eqno(3.11)
$$

\vskip \baselineskip
\noindent
where we have used Eq. (3.10), and $C_{1}$ is a constant resulting from 
the above integration. One notices that the equation above is equivalent to
the one 
obtained from the maximum-entropy principle [cf. Eq. (3.8)]. 
 
An important -- and complementary -- property required for a functional
satisfying the H-theorem is that it should be bounded from below,  
 
$$
F(P(x,t)) \ge F(P_{\rm eq}(x)) \quad (\forall t)~.   
\eqno(3.12)
$$

\vskip \baselineskip
\noindent 
Herein, we assume the presence of a unique equilibrium state in the
functional $F(P(x,t))$. In this case, Eq. (3.12) together with the
imposition from the 
H-theorem, for a time-decreasing functional $F$,   
ensure that, after a long time, the system will always reach equilibrium.
Therefore, it is sufficient to prove that the requirement of Eq. (3.12) holds
only in the nearness of the global equilibrium. Let us then consider,  

$$
F(P)-F(P_{\rm eq}) = \int_{-\infty}^{\infty}dx \ \phi(x) (P - P_{\rm eq})
- {1 \over \beta} \{ \Lambda (Q[P]) -  \Lambda (Q[P_{\rm eq}]) \}~,  
\eqno(3.13)
$$

\vskip \baselineskip
\noindent 
which may be expanded, near the equilibrium, up to   
$O[(P-P_{\rm eq})^{2}]$. It should be noticed that an expansion on the
probability $P(x,t)$, near $P_{\rm eq}(x)$, implies an expansion of the
functional $\Lambda (Q[P])$ in powers of 
$Q[P]-Q[P_{\rm eq}]$; carrying out such an expansion, one gets that

\vspace{-5mm}
\setcounter{enumi}{3}
\setcounter{enumii}{14}
\begin{eqnarray}
F(P)-F(P_{\rm eq}) & = & \int_{-\infty}^{\infty}dx \ \left\{ (P - P_{\rm eq}) 
\left( \phi(x) - {1 \over \beta} 
\left. \deriv{\Lambda[Q]}{Q} \deriv{g[P]}{P}\right|_{P_{\rm eq}(x)} \right) \right. 
\nonumber \\  \nonumber \\
& + & \left. {1 \over 2} (P-P_{\rm eq})^{2}
\left( - {1 \over \beta} 
\left. \deriv{\Lambda[Q]}{Q} \deriv{^2g[P]}{P^2}\right|_{P_{\rm eq}(x)} 
\right) \right\}
\nonumber \\  \nonumber \\
& - & {1 \over 2\beta} 
\left. \deriv{^2\Lambda[Q]}{Q^2}\right|_{P_{\rm eq}(x)} \left( 
\int_{-\infty}^{\infty}dx \ \left. \deriv{g[P]}{P}\right|_{P_{\rm eq}(x)}
(P - P_{\rm eq}) \right)^{2} + \cdots  
\end{eqnarray}

\vskip \baselineskip
\noindent
For the term inside the first integral that appears multiplying 
$(P-P_{\rm eq})$, one may use Eq.
(3.11) in order to get an arbitrary constant; after integration, using the
normalization condition of Eq. (2.9), this first-order term yields zero. For
the term inside the first integral that multiplies $(P-P_{\rm eq})^{2}$, 
one may use Eq. (3.5) at equilibrium, in such a way that, 

\vspace{-5mm}
\setcounter{enumi}{3}
\setcounter{enumii}{15}
\begin{eqnarray}
F(P)-F(P_{\rm eq}) & = & \int_{-\infty}^{\infty}dx \  \left. 
{1 \over 2} (P-P_{\rm eq})^{2} \left\{ \frac{\Omega[P]}{\Psi[P]} \right\}
\right|_{P = P_{\rm eq}(x)}
\nonumber \\  \nonumber \\
& - & {1 \over 2\beta} 
\left. \deriv{^2\Lambda[Q]}{Q^2}\right|_{P_{\rm eq}(x)} \left( 
\int_{-\infty}^{\infty}dx \ \left. \deriv{g[P]}{P}\right|_{P_{\rm eq}(x)}
(P - P_{\rm eq}) \right)^{2} + \cdots  
\end{eqnarray}

\vskip \baselineskip
\noindent
The equation above yields $[F(P)-F(P_{\rm eq})] \geq 0$ provided that one
uses the previously defined properties for the quantities
$\Omega[P]$ and  $\Psi[P]$, and additionally, one supposes that
$(d^{2}\Lambda[Q]/dQ^2)|_{P_{\rm eq}(x)})<0$.

Let us now analyze the situation of an isolated system, i.e., 
$\phi(x)=$ constant; in this case, 
the H-theorem should be expressed in terms of the time derivative of the 
entropy, in such a way that Eq. (3.4) should be replaced by

\vspace{-5mm}
\setcounter{enumi}{3}
\setcounter{enumii}{16}
\begin{eqnarray}
\deriv{S[P]}{t} & = & - \int_{-\infty}^{\infty}dx \ \left( \Omega[P] 
\pderiv{P}{x} \right) 
\left( \deriv{\Lambda[Q]}{Q} \deriv{^2 g[P]}{P^2} \pderiv{P}{x} \right)
\nonumber \\  \nonumber \\
& = & - \int_{-\infty}^{\infty}dx \ \Omega[P] \ 
\deriv{\Lambda[Q]}{Q} \deriv{^2 g[P]}{P^2} 
\left( \pderiv{P}{x} \right)^{2}\geq 0~. 
\end{eqnarray}

\vskip \baselineskip
\noindent
As expected, the proof of the H-theorem for an isolated system becomes much
simpler than  
that for the system in the presence of an external potential. In
particular, there is no requirement for a relation involving the
parameters of the FPE and the entropy, like the one of Eq. (3.5); all that
one needs is a standard condition associated with the FPE, i.e., 
$\Omega[P] \geq 0$,  the restriction $d \Lambda [Q] / d Q \geq 0$
for the outer functional of the entropy, as well as the general
restrictions of Eq. (3.1) for the entropy. 

\vskip 2 \baselineskip
\noindent
{\large\bf 4. \quad Some Families of FPEs and their Associated Entropies}

\vskip \baselineskip

In this section we will explore further the correspondence 
between the nonlinear FPE
of Eq. (2.5) and general entropic forms, established through Eq. (3.5). This
equation shows clearly that there may be families of FPEs, 
corresponding to the same ratio  
$(\Omega[P]/\Psi[P])$, associated with a single entropic form, i.e.,  
the same entropy may be associated with different dynamical processes. 
In the following examples, we consider classes of FPEs satisfying 

$$
\Omega[P] = a[P]b[P]~; \quad \Psi[P] = a[P]P~, 
\eqno(4.1)
$$

\vskip \baselineskip
\noindent
where the functionals $a[P]$ and $b[P]$ are restricted by the conditions
imposed previously for the functionals  
$\Omega[P]$ and $\Psi[P]$. In addition to that, in the first three examples
we will consider entropic forms characterized by $\Lambda (Q[P]) = Q[P]$;
for these cases Eq. (3.5) becomes

$$
\deriv{^2 g[P]}{P^{2}} = - \beta \ \frac{b[P]}{P}~. 
\eqno(4.2)
$$

\vskip \baselineskip
\noindent
Therefore one has a freedom for choosing different forms for the 
functional $a[P]$, leading to the same entropic form.
Next, we work out some examples.  

\vskip \baselineskip

a) {\it The class of FPEs associated with the Boltzmann-Gibbs entropy}:
this class corresponds to the functionals $\Omega[P]$ and $\Psi[P]$
satisfying Eq. (4.1), with  
$b[P]=D$ (constant). Integrating Eq. (4.2) one gets,  

$$
\deriv{g}{P} = -\beta D \ln P + C  \quad \Rightarrow \quad 
g[P] = -k_{B}P \ln P~, 
\eqno(4.3)
$$

\vskip \baselineskip
\noindent
where we have used the conditions \ $g(0)=g(1)=0$ \ to eliminate the 
constant $C$, 
and set the Lagrange multiplier $\beta=k_{B}/D$, where $k_{B}$ represents
the Boltzmann constant. It should be stressed that, usually one associates
the Boltzmann-Gibbs entropy with the linear FPE, which represents the
simplest equation within the present class. Herein we show that, by
properly defining the functionals $\Omega[P]$ and $\Psi[P]$, one may get
nonlinear FPEs, with time-dependent solutions that may be different from
standard exponential probability
distributions, but still associated with the Boltzmann-Gibbs entropy. This
whole family of FPEs presents the Boltzmann-Gibbs distribution as the
stationary-state solution. 
As a simple example of this class, one may have the nonlinear FPE
characterized by  
$a[P]=P^{\nu}$ ($\nu \in \Re$) and $b[P]=D$ (constant). 

\vskip \baselineskip 

b) {\it The class of FPEs associated with Tsallis' entropy}: It is
important to notice  
that the simplest FPE of this class was originally proposed with 
$\Psi[P(x,t)]=P(x,t)$ and  
$\Omega [P(x,t)]=(2-q)D[P(x,t)]^{1-q}$, where $D$ is a constant 
\cite{plastino95}; however, it is very common in the literature 
\cite{next99,next04,next05} to find this FPE with the replacement 
$2-q \rightarrow q$. Herein we shall consider this class of FPEs in such a
way to satisfy Eq. (4.1), with $b[P(x,t)]=qD[P(x,t)]^{q-1}$;  integrating
Eq. (4.2), 

$$
g[P] = - \frac{\beta D}{q-1} P^{q} + CP \quad \Rightarrow \quad 
g[P] = k \frac{P-P^{q}}{q-1}~, 
\eqno(4.4)
$$

\vskip \baselineskip
\noindent
where we have set $\beta=k/D$ ($k$ is a constant with dimensions of entropy) 
and have also used the conditions \ $g(0)=g(1)=0$ \ to eliminate 
the constant $C$. In Eq.
(4.4) one readily recognizes the entropy proposed by Tsallis
\cite{tsallis88}, that depends on the well-known entropic index $q$.
Similarly to  example (a), one has a whole class of FPEs, corresponding to
different choices for the functional  
$a[P]$ of Eq. (4.1), some of which exhibit time-dependent
solutions different from the ones presented in Refs. 
\cite{plastino95,tsallis96}, but all of them associated with the
entropic form of Eq.~(4.4). This whole family of FPEs presents the Tsallis
distribution (also known as $q$-exponential) \cite{next99,next04,next05} as
the stationary-state solution. 

\vskip \baselineskip

c) {\it The class of FPEs associated with the entropy of Refs. 
\cite{curadobjp,curado04}}: In this example we proceed in an inverse way
with respect to the previous two cases, i.e., we start from a given
entropic form, in order to find the class of FPEs associated with it. Let
us then consider \cite{curadobjp,curado04}, 

$$
g[P] = k [1 - \exp(-cP) + Pg_{0}]~; \quad (g_{0}=\exp(-c)-1)~,
\eqno(4.5)
$$

\vskip \baselineskip
\noindent
where $c$ is an arbitrary dimensionless constant, and $k$ is a constant
with dimensions of entropy.  Substituting into Eq. (4.2), one gets 

$$
b[P] = - DP[1 - c^{2}\exp (-cP)]~,
\eqno(4.6)
$$

\vskip \baselineskip
\noindent
where we have set $D=k/\beta$. 
The functional form above defines the family of
FPEs associated with different definitions for the functional $a[P]$, 
all to them related to the entropic form of Eq. (4.5); the simplest of
these equations corresponds to $a[P]=1$. 

\vskip \baselineskip

d) {\it The class of FPEs associated with the Renyi entropy \cite{renyi}}: 
Similarly to the previous example, we start from the entropic form, in
order to find the class of FPEs associated with it. In this case we have
that  

$$
\Lambda (Q[P]) = k \ {\ln Q[P] \over 1-q}~;  \quad 
\deriv{\Lambda[Q]}{Q} = {k \over (1-q)Q[P]}~; \quad g[P] = P^{q}~, 
\eqno(4.7)
$$

\vskip \baselineskip
\noindent
where $k$ is a constant with dimensions of entropy. It is important to
stress that in order to satisfy the H-theorem, entropic forms characterized
by an outer functional  
$\Lambda[Q]$ are restricted to the condition that
$(d \Lambda[Q] / dQ)$ should present a sign different from
the one of $(d^{2}g[P]/dP^{2})$ (like assumed in the beginning of section
3), as can be seen from simple 
analyses of Eq. (3.5), for the case of a system in the presence of an
external potential, or of Eq. (3.16), for the case of an isolated system.
Substituting the functionals of Eq.~(4.1) into Eq.~(3.5) one gets,  

$$
- \frac{1}{\beta} \deriv{\Lambda[Q]}{Q} \deriv{^2 g[P]}{P^{2}} =  
\frac{b[P]}{P}~,   
 \eqno(4.8)
$$

\vskip \baselineskip
\noindent
and using Eq.~(4.7), 

$$
b[P] =  \frac{Dq}{Q[P]} P^{q-1} =  
\frac{DqP^{q-1}}{\int_{-\infty}^{\infty}dx \ P^{q}}~,  
\eqno(4.9)
$$

\vskip \baselineskip
\noindent
where we have set $D=k/\beta$. It is important to remind that the functionals 
$\Omega[P]$ and $\Psi[P]$ are supposed to be both positive,
for a well-defined FPE, which imply $a[P],b[P]>0$ [cf. Eq.~(4.1)].
>From Eq.~(4.9) this condition is not satisfied if $q \leq 0$. 
Notice that this entropic form satisfies the condition
$(d^{2}\Lambda[Q]/dQ^2)|_{P_{\rm eq}(x)})<0$, required by the H-theorem
[cf. Eq. (3.15)], for $q<1$; therefore, one can assure the validity of such
an entropic form, from the physical point of view, for the interval $0<q<1$.

\vskip 2 \baselineskip
\noindent
{\large\bf 5. \quad A FPE for a More General Free-Energy Functional}

\vskip \baselineskip

In this section we will consider a slightly different FPE, with respect to
the one of Eq. (2.5), namely, 

$$
\pderiv{P(x,t)}{t} = \pderiv{}{x} \left( \Psi[P] \pderiv{}{x} 
\left( \phi(x) \chi[P]\right)
\right) + \pderiv{}{x} \left( \Omega[P] \pderiv{P}{x} \right)~, 
\eqno(5.1)
$$

\vskip \baselineskip
\noindent
where a new functional $\chi[P]$ was introduced [notice that Eq. (2.5) is
recovered for $\chi[P]=1$], that should be finite and positive definite. 
The interesting point about such a FPE is that it is
consistent with the definition of a ``generalized internal energy''
\cite{next99,next04,next05}, 

$$
U =  \int_{-\infty}^{\infty}dx \ \phi(x) \Gamma [P(x,t)]~,  
\eqno(5.2)
$$

\vskip \baselineskip
\noindent
where we are assuming that $\Gamma[P]$ represents a positive, monotonically
increasing  
functional of $P(x,t)$, that should be at least $\Gamma[P] \in C^{1}$. 

Now, we take this internal energy in the free-energy
functional of Eq. (3.2) and consider the same entropic form of Eq. (3.1).  
Let us then prove the H-theorem for such a system, following the same
steps of Section 3; one gets that, 

\vspace{-5mm}
\setcounter{enumi}{5}
\setcounter{enumii}{3}
\renewcommand{\theequation}{\arabic{enumi}.\arabic{enumii}}
\begin{eqnarray}
\deriv{F}{t} & = & \deriv{}{t} \left(\int_{-\infty}^{\infty}dx \ 
\phi(x) \Gamma [P(x,t)]
- \frac{1}{\beta} \Lambda(Q[P])  \right)  \nonumber \\ 
\nonumber \\
& = & \int_{-\infty}^{\infty}dx \ \left( \phi(x) \deriv{\Gamma[P]}{P} 
- \frac{1}{\beta} \deriv{\Lambda[Q]}{Q} \deriv{g[P]}{P} \right) 
\pderiv{P}{t}~.    
\end{eqnarray}

\vskip \baselineskip
\noindent
Using the FPE of Eq.~(5.1) and integrating by parts, one obtains

\vspace{-5mm}
\setcounter{enumi}{5}
\setcounter{enumii}{4}
\begin{eqnarray}
\deriv{F}{t} & = & - \int_{-\infty}^{\infty}dx \ 
\left\{ \Psi[P] \pderiv{}{x} \left(
\phi(x) \chi[P]\right) + \Omega[P] \pderiv{P}{x} \right\} \nonumber \\
\nonumber \\
& \times &
\left\{ \pderiv{}{x} \left(  \phi(x) \deriv{\Gamma[P]}{P}\right)
-\frac{1}{\beta} \deriv{\Lambda[Q]}{Q} 
\deriv{^2 g[P]}{P^2} \pderiv{P}{x} \right\}~.  
\end{eqnarray}

\vskip \baselineskip
\noindent
The H-theorem applies, i.e., 

$$
\deriv{F}{t} = - \int_{-\infty}^{\infty}dx \ 
\Psi[P] \Bigl\{ \pderiv{}{x} \left( \phi(x) 
\chi[P] \right) +
  \frac{\Omega[P]}{\Psi[P]} \pderiv{P}{x} \Bigr\}^2 \leq 0~,
\eqno(5.5)
$$

\vskip \baselineskip
\noindent
provided that Eq. (3.5) holds, with an additional restriction for the
functional $\chi[P]$, 

$$
\chi[P] = \deriv{\Gamma[P]}{P}~.  
\eqno(5.6)
$$

\vskip \baselineskip
\noindent
It should be mentioned that the constraint above, relating the functional
$\chi[P]$ of the FPE with 
the quantity $\Gamma[P]$ that appears in the definition of the 
generalized internal energy (with $\Gamma[P] \ne P$), has to be introduced,
in such a way to satisfy the H-theorem. 

Let us now show that, at equilibrium, the condition of Eq.~(3.5) is
equivalent to the maximum-entropy principle, when one uses the FPE of 
Eq.~(5.1). Defining the functional  

$$
{\cal I}[P(x,t)] = \Lambda (Q[P])
+ \alpha \left( 1 - \int_{-\infty}^{\infty}dx \ P(x,t) \right)
+ \beta \left(U - \int_{-\infty}^{\infty}dx \ \phi(x) \Gamma [P(x,t)] \right)~,
\eqno(5.7)
$$

\vskip \baselineskip
\noindent 
($\alpha$ and $\beta$ are Lagrange multipliers) one has that, 

$$
\left. \deriv{{\cal I}[P]}{P} 
\right|_{P = P_{\rm eq}(x)}= 0 \quad \Rightarrow \quad 
\left.  \deriv{\Lambda[Q]}{Q} \deriv{g[P]}{P}\right|_{P = P_{\rm eq}(x)} 
- \alpha - \beta \, \phi(x) 
\left. \deriv{\Gamma[P]}{P} \right|_{P = P_{\rm eq}(x)} = 0~,  
\eqno(5.8)
$$

\vskip \baselineskip
\noindent 
where $P_{\rm eq}(x)$ represents the probability distribution at
equilibrium. Considering Eq.~(5.1) at equilibrium one gets, 

$$
-\pderiv{}{x} \left( \phi(x) \chi[P_{\rm eq}]\right)
= {\Omega[P_{\rm eq}] \over \Psi[P_{\rm eq}]} 
\deriv{P_{\rm eq}(x)}{x}~,  
\eqno(5.9)
$$

\vskip \baselineskip
\noindent
and after integration,

$$
-\phi(x) \chi[P_{\rm eq}(x)] + C = \int_{x_{0}}^{x}dx \  
{\Omega[P_{\rm eq}] \over \Psi[P_{\rm eq}]} 
\deriv{P_{\rm eq}(x)}{x}
= \int_{P_{\rm eq}(x_{0})}^{P_{\rm eq}(x)} \ 
{\Omega[P_{\rm eq}(x^{\prime})] \over 
\Psi[P_{\rm eq}(x^{\prime})]} 
dP_{\rm eq}(x^{\prime})~,  
\eqno(5.10)
$$

\vskip \baselineskip
\noindent
where $C \equiv \phi(x_{0})\chi[P_{\rm eq}(x_{0})]$ is a constant. 
Integrating Eq. (3.5), at equilibrium, and using the equation above, one
gets,  

$$
\frac{1}{\beta} \left. \deriv{\Lambda[Q]}{Q} 
\deriv{g[P]}{P}\right|_{P_{\rm eq}(x)} 
= \phi(x) \chi[P_{\rm eq}(x)] + C^{\prime} 
= \left. \phi(x) \deriv{\Gamma[P]}{P} \right|_{P = P_{\rm eq}(x)} 
+ C^{\prime}~,   
\eqno(5.11)
$$

\vskip \baselineskip
\noindent
where we have used Eq.~(5.6) and $C^{\prime}$ represents another integration
constant.  
The equation above is equivalent to Eq. (5.8), 
obtained from the maximum-entropy principle.

Therefore, in what concerns the H-theorem, the necessary relation involving
quantities of the FPE with a general entropic form and its equivalence with
the maximum-entropy principle, the FPE of Eq. (5.1) is consistent with the
definition of a generalized internal energy that is sometimes used in the
context of nonextensive statistical mechanics \cite{next99,next04,next05}. 

\vskip 2 \baselineskip
\noindent
{\large\bf 6. \quad Conclusions}

\vskip \baselineskip

We have proved the H-theorem by using general nonlinear
Fokker-Planck equations. In order to prove the H-theorem for a system in 
the presence of an external
potential, a relation involving terms of the Fokker-Planck
equation and the entropy of the system was proposed. 
In principle, one may have classes of Fokker-Planck equations related to a 
single entropic form. 
Since the Fokker-Planck equation is a phenomenological equation that
specifies the dynamical evolution associated with a given physical system, 
this relation may be useful in the identification of  
the entropic form associated with such a system.
In particular, the present approach makes it possible to identify 
entropic forms 
associated with some anomalous systems, exhibiting unusual behavior, that
are known to be appropriately described by nonlinear Fokker-Planck
equations, like the ones considered herein.
By considering a modified Fokker-Planck equation, we have also proved the
H-theorem for a type of generalized internal energy,
like the one used within the nonextensive statistical-mechanics formalism.   
For that, the same relation connecting the parameters of the
Fokker-Planck equation and the corresponding entropic form had to be introduced.
To our knowledge, it is first time that the H-theorem has been verified, 
for a system in the presence of an external potential, by
considering a nonlinear weight in the definition of the internal energy. 
Making use of the relation mentioned, we have calculated well-known entropic
forms, associated with given Fokker-Planck equations. In the case of the
standard Boltzmann-Gibbs 
entropy, apart from the simplest, linear Fokker-Planck equation, one may
have a whole class of nonlinear Fokker-Planck equations, whose
time-dependent probability
distributions may be distinct from simple exponential distributions, 
but all of them 
related to this particular entropic form; the stationary-state solution is
the same as the one of the linear Fokker-Planck equation, i.e., a
Boltzmann-Gibbs distribution. A similar behavior is verified
for more general, nonadditive, entropic forms, e.g., the Tsallis' entropy.
Although this relation involving families of Fokker-Planck equations and 
entropic forms 
may not be unique, we have shown that, when considered at equilibrium, it
is equivalent to the principle of maximum entropy. 
The present results suggest that behind such a relation there may be a
deep physical insight that deserve further investigations. 

\vskip 2\baselineskip

{\large\bf Acknowledgments}

\vskip \baselineskip
\noindent
We thank Profs. Angel Plastino and Constantino Tsallis for fruitful
conversations. The partial financial supports from
CNPq and Pronex/MCT/FAPERJ (Brazilian agencies) are acknowledged. 

\vskip 2\baselineskip

\end{document}